\documentclass[12pt,preprint2]{aastex}
\usepackage{emulateapj5}
\usepackage{onecolfloat5}
\usepackage{natbib}
\newcommand{\rxj}{RX~J1605.3+3249}
\newcommand{\rxjw}{RX~J1856.5$-$3754}
\newcommand{\rbs}{RX~J1308.6+2127}
\newcommand{\rxjk}{RX~J0720.4$-$3125}
\newcommand{\chandra}{\textit{Chandra}}

\newcommand{\hst}{\textit{HST}}
\newcommand{\rosat}{\textit{ROSAT}}
\newcommand{\xmm}{\textit{XMM}}

\newcommand{\beq}{\begin{equation}}
\newcommand{\eeq}{\end{equation}}

\newcommand{\expnt}[2]{\ensuremath{#1 \times 10^{#2}}}   

\shortauthors{Kaplan et al.}
\shorttitle{The Optical Counterpart of \rxj}

\slugcomment{Accepted by ApJL}
\begin{document}
\twocolumn[
\title{The Optical Counterpart of the Isolated Neutron Star \rxj}
\author{D.~L.~Kaplan, S.~R.~Kulkarni}
\affil{Department of Astronomy, 105-24
California Institute of Technology, Pasadena, CA 91125, USA}
\email{dlk@astro.caltech.edu, srk@astro.caltech.edu} 
\and 
\author{M.~H.~van~Kerkwijk}
\affil{Dept. of Astronomy \& Astrophysics, University of
  Toronto, 60 St George St.,
  Toronto, ON, M5S 3H8, Canada}
\email{mhvk@astro.utoronto.ca}

\begin{abstract}
We have detected the optical counterpart to the nearby isolated
neutron star \rxj\ using observations from the Space Telescope Imaging
Spectrometer aboard the \textit{Hubble Space Telescope}.  The
counterpart, with $m_{\rm 50CCD}=26.84\pm0.07$~mag and very blue
colors, lies  close to the \rosat\ HRI error circle and within
the \chandra\ error circle.  The spectrum is consistent with a
Rayleigh-Jeans tail whose emission is a factor of $\approx 14$ above
the extrapolation of the X-ray blackbody, and the source has an
unabsorbed X-ray-to-optical flux ratio of $\log(f_{X}/f_{\rm
opt})=4.4$, similar to that of other isolated neutron stars.  This
confirms the classification of \rxj\ as a neutron star.
\end{abstract}

\keywords{pulsars: individual (\rxj)---stars: neutron---X-rays: stars}
]
\section{Introduction}

Thanks to {\em ROSAT}, a half dozen nearby neutron stars that emit no
detectable radio emission have been
identified (see \citealt{ttzc00} for a
comprehensive review).  These objects have been eagerly studied using  facilities such as the \textit{XMM-Newton}
Observatory, the \textit{Chandra} X-ray Observatory, the
\textit{Hubble Space Telescope} (\hst), Keck and the Very Large
Telescope with the hope of
determining the physical parameters, in particular radius and
temperature, and to compare these to models of neutron
stars.  
Independently, by their sheer proximity these objects  play a
pivotal role in assessing the neutron star demographics in the
Galaxy. These two considerations highlight the virtue of detailed
studies of nearby neutron stars.

\rxj\ was identified in the \rosat\ All-Sky Survey by
\cite{mhz+99}. The X-ray spectrum is well fitted by a blackbody with
$kT \sim 90\,$eV and low interstellar column density, $N_H \sim
10^{20}\,$cm$^{-2}$, and is quite similar to those of well studied
nearby neutron stars such as \rxjw\ and \rxjk.  \cite{mhz+99} obtained
deep ($B\sim 27\,$mag and $R\sim 26\,$mag) images from the Keck
telescope. Only one object (star C) was found within the $2\arcsec$
High Resolution Imager (HRI) circle (we believe that the uncertainty
of the HRI position was underestimated; see \S~\ref{sec:opt}).
Optical spectroscopic observations carried out at the
Canada-France-Hawaii Telescope (CFHT) showed that star C was a distant
late-type M dwarf.

However, the soft spectrum and the stable X-ray emission are better
accounted for by a model in which \rxj\ is an isolated neutron
star. If so, the optical counterpart would be below (or perhaps just
at) the limit of the Keck observations.  As a part of our
investigation of nearby neutron stars with \hst\ we undertook deep
observations of this field.  In this Letter we report the discovery of
a faint blue optical star which we identify with the optical
counterpart of \rxj. Our discovery confirms that \rxj\ is a nearby
neutron star.

\section{Observations \& Data Reduction}

\begin{deluxetable}{c c c c c}
\tablecaption{Summary of Optical Observations\label{tab:obs}}
\tablewidth{0pt}
\tablehead{
\colhead{Telescope} & \colhead{Instrument} & \colhead{Date} &
\colhead{Exposure} & \colhead{Band} \\
 & & \colhead{(UT)} & \colhead{(s)} \\
}
\startdata
Keck~II & LRIS & 1998-Aug-24 & 1800 & R \\
\hst & STIS & 2001-Jul-21 & 2700 & 50CCD \\
\hst & STIS & 2001-Jul-21 & 5360 & F28X50LP \\
\enddata
\end{deluxetable}

\subsection{Hubble Space Telescope Observations}

We observed \rxj\ with the Space Telescope Imaging Spectrograph (STIS)
aboard \hst\ in two modes:
unfiltered CCD (50CCD aperture) and a longpass filter that transmitted
photons longward of $\approx 5500$~\AA; see 
Table~\ref{tab:obs}.  For each mode,
the individual images were drizzled \citep{fh02} onto
a single image with a pixel scale of 0.5. Thus final images 
had $0\farcs0254$ pixels.

\subsection{Keck Observations}
\label{sec:opt}
\begin{figure}[t]
\plotone{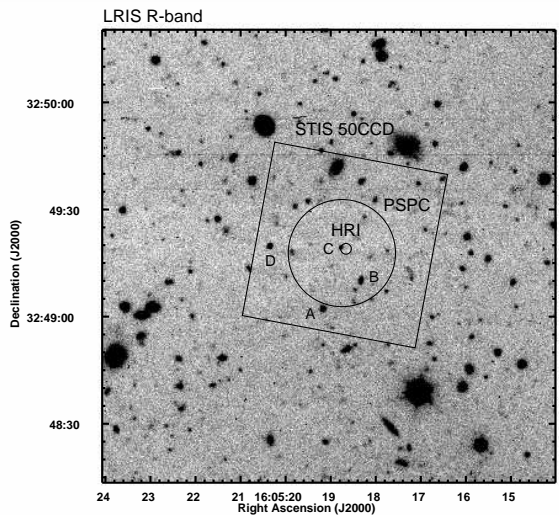}
\caption{LRIS R-band image of the field around \rxj.  The
   field-of-view of the STIS 50CCD observation is 
   indicated by the box.  The $15\arcsec$-radius (90\% confidence)
   PSPC and $1\farcs5$-radius (90\% confidence) updated HRI error circles are
   also indicated.  Sources A, B, C, and D from \citet{mhz+99} are
   labeled.
\label{fig:lris}
}
\end{figure}

For astrometric purposes we obtained imaging data from  
the Low-Resolution Imaging Spectrometer (LRIS;
\citealt{o+95}) on the Keck~II telescope; see Table~\ref{tab:obs}.  
These images were reduced in a standard manner using tasks in MIDAS:
bias subtraction, flat-fielding, and stacking of the
exposures.  We show the LRIS image in Figure~\ref{fig:lris}.

After correcting the stellar positions measured in a 30-s LRIS image
for geometric distortions\footnote{See
\url{http://alamoana.keck.hawaii.edu/inst/lris/coordinates.html}.}, we
fit for plate-scale, zero-point, and rotation using 20 unsaturated
stars from the latest version of the Guide Star Catalog
(GSC-2.2)\footnote{See
\url{http://www-gsss.stsci.edu/support/data\_access.htm}.}, obtaining
rms residuals of $0\farcs09$ in each coordinate (in what follows,
astrometric uncertainties refer to rms values in each coordinate
unless otherwise specified).  We then determined the astrometric
solution (plate-scale, zero-point, and rotation) of the full 30-minute
LRIS image by using 24 stars common to that and the 30-s image,
getting rms residuals of $0\farcs02$.  Finally, we used 21 stars on
the deep LRIS image to determine the plate-scale, zero-point, and
rotation of the drizzled STIS 50CCD image (shown in
Fig.~\ref{fig:stis}),  getting rms residuals of
$0\farcs04$.  The final uncertainty to which our STIS coordinates are
on the ICRS is dominated by the $0\farcs3$ uncertainty of the
GSC-2.2\footnote{See
\url{http://www-gsss.stsci.edu/gsc/gsc2/calibrations/astrometry/astrometry.htm\#method}.}.

We inspected all point sources on the STIS images that were within the
PSPC error circle and found a very blue source. As can be seen from
Figure~\ref{fig:cm} this is the bluest source in the PSPC error
circle.  The source, hereafter ``X'', is located at (J2000)
$\alpha=16^{\rm h}05^{\rm m}18\fs52$
$\delta=+32\degr49\arcmin18\farcs0$, with uncertainties of about
$0\farcs3$, and lies $2\farcs5$ from the HRI position from
\citet{mhz+99}, outside the nominal 90\% error circle.  

However, we have found a problem in comparing positions referenced to
the USNO-A2.0 \citep{m98}, like the HRI position, and those referenced
to the GSC-2.2/ICRS (our optical data).  In this field, there appears
to be a systematic shift between the GSC-2.2 and the USNO-A2.0 of
$\langle \delta_{\rm USNO}-\delta_{\rm GSC}\rangle=0\farcs65$ (the
shift in Right Ascension is a negligible $0\farcs02$).  If we correct
the HRI position to the GSC-2.2 reference frame, we find the new
position to be (J2000) $\alpha=16^{\rm h}05^{\rm m}18\fs66$
$\delta=+32\degr49\arcmin19\farcs0$.  With this position, the error
circle appears to be located properly with respect to star C,
comparing with Fig.~4 of \citet{mhz+99}.

Even with the updated position, X is slightly outside the 90\% HRI
error circle\footnote{X is within the error circle of the position
obtained from a preliminary analysis of \xmm\ data; van Kerkwijk et
al., in preparation.}.  We have two possible explanations for this.
First, \rxj\ may have non-negligible proper motion, such as that seen
for \rxjw\ \citep{wal01}.  In this case the offset between the HRI
position (epoch 1998.3) and the STIS image (2001.6) could be real.
However, the LRIS data are not of sufficient quality to detect X with
any confidence, so we will have to wait for additional data.  Second,
we note that the quoted uncertainty of the HRI position may be
underestimated: \citet{mhz+99} used 6 reference sources for the
boresight corrections and claim an uncertainty of $0\farcs64$ with no
contribution from systematic effects.  In comparison, \citet{hbg+98}
use 32 sources and get typical HRI uncertainties of $1.0\arcsec$ that
include a $0\farcs5$ systematic error to achieve good X-ray-to-optical
matches.

In either case, the blue color of source X is similar to those of the
counterparts of other isolated neutron stars (e.g.\ \rxjw\ and \rxjk;
\citealt{wm97,vkk01,mh98,kvk98}).  Thus we consider it likely that X
is the counterpart of \rxj.

\subsection{X-ray}
While our identification based on color and position is plausible, the
uncertainty in the \rosat\ HRI position prevents us from being sure
about the association (pulsations or a common proper motion would be
definitive measurements).
Fortunately, the availability of archival data from the 
\textit{Chandra X-ray Observatory} offered us to opportunity
to decrease the chance coincidence probability by a factor of 10.
The {\em Chandra}
observation (ObsID 2791) had a duration of 20-ks, and \rxj\ was at
the aim-point of the ACIS-I CCD array.  
Using standard processing 
steps\footnote{\url{http://asc.harvard.edu/cal/ASPECT/fix\_offset/fix\_offset.cgi}.}
we corrected for a systematic astrometric error of $\Delta
\alpha=-0\farcs35$ and $\Delta \delta=-0\farcs10$.  
As a cross-check, we compared the positions of other X-ray sources
to GSC-2.2 stars (which, since we used the GSC-2.2 as the reference
for the optical astrometry, ensures that they are on the same system
as our \hst\ data) and found that the coordinates match to better than
0.5 arcsecond.

\begin{figure}[t]
\plotone{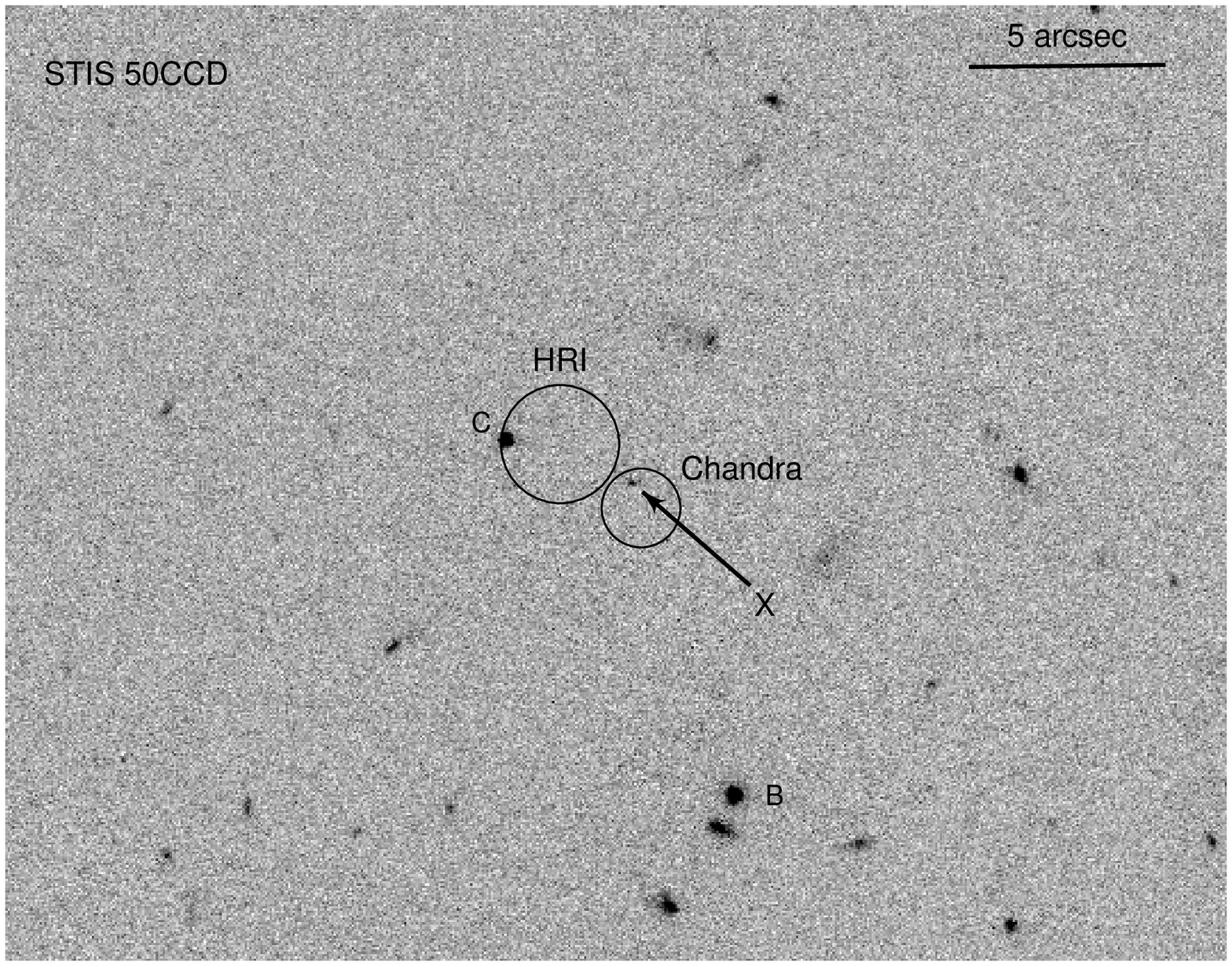}
\caption{STIS/50CCD image of \rxj.  The image has North up and East to
  the left. A $5\arcsec$ scale bar is in the upper right.  The
  updated 90\% confidence error circles from
  \rosat/HRI  ($1\farcs5$ radius) and  \chandra\ 
  ($1\farcs0$ radius) are plotted.  The likely counterpart to \rxj\ is
  indicated by the X, and  sources B and C from \citet{mhz+99} are labeled
  (source B is not just extended to the South-East, as noted by
  \citealt{mhz+99}, but is in fact composed of separate sources).  The
  PSPC error circle (Fig.~\ref{fig:lris}) is larger than this image.
}
\label{fig:stis}
\end{figure}

\begin{figure}
\plotone{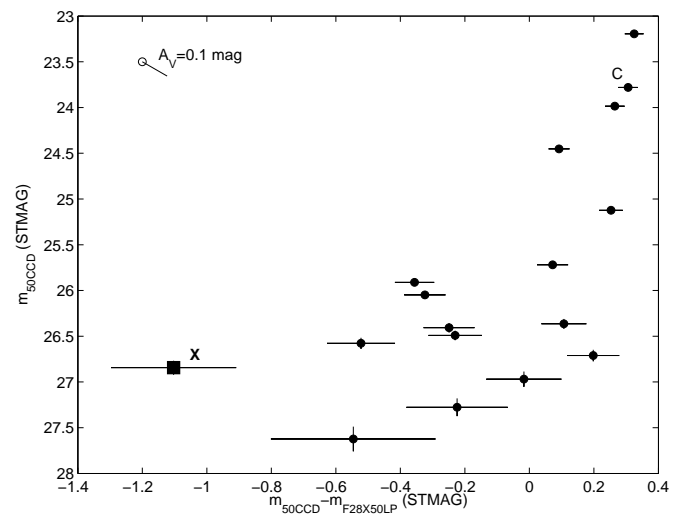}
\caption{Color-magnitude diagram of point-like source from the STIS
  data.  50CCD magnitude is plotted against ${\rm 50CCD}-{\rm
  F28X50LP}$ color.  Source X is indicated and plotted as a square.
  Source C is also indicated.  The sources have been corrected to
  ``infinite'' aperture using the correction most appropriate for \rxj\
  (\S~\ref{sec:ana}).  A reddening vector is plotted for
  $A_{V}=0.1$~mag.}
\label{fig:cm}
\end{figure}

We then measured the centroid of \rxj\ (with a count-rate of $\approx
0.15\mbox{ s}^{-1}$, \rxj\ is somewhat affected by photon pileup, but
this should not affect the centroid) to be (J2000) $\alpha=16^{\rm
  h}05^{\rm m}18\fs50$, $\delta=+32\degr 49\arcmin 17\farcs4$.  We
estimate a final 90\% confidence radius of the X-ray position with
respect to the STIS image of $\approx 1\farcs0$.  As can be seen in
Fig.~\ref{fig:stis}, source X is well within this radius, lending
credence to our identification.

\section{Analysis \& Discussion}
\label{sec:ana}
We rely on the spectral fits to the \rosat\ PSPC data
presented in \citet{mhz+99}: $kT=92$~eV, $N_{H}=\expnt{1.1}{20}\mbox{
  cm}^{-2}$, and $R_{\infty}=3.3 d_{300}$, where $d=300 d_{300}$~pc is the
distance and the normalization assumes $0.9\mbox{ counts s}^{-1}$ in
the PSPC.  The absorption column density implies an extinction of 
$A_{V}=0.06$~mag, \citep{ps95} --- while
this value is uncertain, it is low enough to not make a large
difference.  The total Galactic hydrogen column density is
$\expnt{2.4}{20}\mbox{ cm}^{-2}$ (determined by COLDEN\footnote{See
  \url{http://asc.harvard.edu/toolkit/colden.jsp}.}; \citealt{dl90})  so that the
maximum extinction is $0.13$~mag.  This agrees with the 
extinction estimated from infrared dust emission \citep*{sfd98},
and confirms that the total extinction to \rxj\ is low.

Within a $0\farcs25$-radius aperture, we measure magnitudes of $m_{\rm
50CCD}=27.03\pm 0.07$~mag and $m_{\rm F28X50LP}=28.16\pm 0.18$~mag for
source X in the STMAG system.  To correct the photometry to a nominal
infinite aperture, we follow \citet*{kkvk02} and \citet{kvkm+03}.  As X
is bluer than all of the stars in the image, we used the bluest of the
available aperture corrections: 0.183~mag at $0\farcs25$ radius for
50CCD and 0.214~mag at $0\farcs25$ radius for F28X50LP (T. Brown 2002,
private communication).  This then gives corrected magnitudes of
$m_{\rm 50CCD}=26.84 \pm 0.07$~mag and $m_{\rm F28X50LP}=27.95 \pm
0.18$~mag, where we have incorporated a 0.02~mag uncertainty from the
aperture corrections.

Since the STIS bandpasses are so wide we must use the shapes of the
bandpasses to convert the measured magnitudes into fluxes.  Following
\citet{vkk01}, we find effective wavelengths $\langle \lambda \rangle$
of 5148~\AA\ and 7137~\AA\ and effective extinctions $\langle
A_{\lambda}/A_{V}\rangle$ of 1.56 and 0.79 for 50CCD and F28X50LP,
respectively.  With these wavelengths, we can now apply the standard
STMAG conversion of $F_{\lambda}(\langle \lambda
\rangle)=10^{-(m+21.1)/2.5}\mbox{ ergs s}^{-1}\mbox{ cm}^{-2}\mbox{
\AA}^{-1}$.

\begin{figure}[t]
\plotone{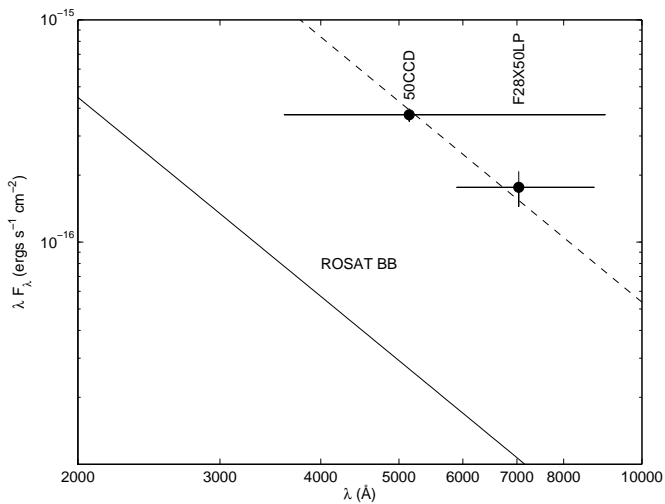}
\caption{Spectral energy distribution for source X corrected for
  absorption with $A_{V}=0.06$~mag.  The STIS data are
  plotted as points.  The extrapolation of the \rosat\ blackbody
  \citep{mhz+99} is the solid line (labeled ``ROSAT BB''), and a Rayleigh-Jeans fit to the
  STIS data is the dashed line.  The horizontal error-bars show the
  bandpasses of the filters.}
\label{fig:nufnu}
\end{figure}

We plot the spectral energy distribution of source X in
Figure~\ref{fig:nufnu}, assuming a blackbody spectrum in the X-ray
regime\footnote{This blackbody is likely a simplification of a more
realistic atmosphere model, but as yet such models have been
unsuccessful in fitting sources like \rxjw\ and \rxjk.}.  As one can see, the
optical photometry appear to follow a power-law with spectral index
$\alpha\approx 2$ ($F_{\nu}\propto \nu^{\alpha}$), appropriate for a
Rayleigh-Jeans tail and similar to that of other isolated neutron
stars \citep{vkk01,kvkm+03}. We also plot in Figure~\ref{fig:nufnu}
the extrapolation of the best-fit \rosat\ PSPC blackbody.  The
Rayleigh-Jeans fit to the STIS data has a normalization that is a
factor of $14\pm 2$ above the extrapolation of the blackbody.  Other
power-law indices are possible, as found for \rxjk\ \citep{kvkm+03},
but we do not believe that the current data warrant a full fit.  The
unabsorbed bolometric X-ray-to-optical flux ratio is
$\log(f_{X}/f_{\rm opt})=4.4$ (assuming that the X-ray spectrum is
well described by a blackbody).

The blueness of source X and its X-ray-to-optical flux ratio, taken
together with the X-ray spectrum of \rxj, virtually guarantee that X
is the optical counterpart of \rxj\ and that \rxj\ is an isolated
neutron star.  It then joins 3 other sources (\rxjw, \rxjk, and
possibly \rbs) that have soft X-ray blackbodies, blue optical
counterparts, and no other emission; see Table~\ref{tab:ins}.  \rxj\
stands out from the other sources in Table~\ref{tab:ins} by virtue of
its relatively large optical excess: the optical flux is a factor of
14 above the extrapolated X-ray flux, where for the other sources the
ratio is closer to 6 (this is despite the fact that the
X-ray-to-optical flux ratio for \rxj\ is within the range of the other
sources).  This can also be seen from the \rosat\ data in
\citet{shs+99}, where \rxj\ has a factor of 2 smaller count rate than
\rxjk\ despite being hotter and having a comparable optical magnitude.
There are two possibilities to explain the large excess of \rxj:
either its X-ray emission is suppressed relative to the optical, or
the optical emission is enhanced.  

The first scenario implies that the blackbody radius ($3.3
d_{300}$~km, where the distance of $\sim 300$~pc is predicted based on
the optical flux; see \citealt*{kvka02}) is significantly smaller than
those of the other sources (typically 6~km).  If the blackbody radius
can be interpreted as the radius of a hot polar cap, perhaps \rxj\ has
a different magnetic field configuration leading to a smaller cap
size, or \rxj\ is in an orientation where only half of the cap is
visible (although this is difficult when relativistic beaming is taken
into account; \citealt*{pod00}).

\begin{deluxetable}{c c c c c c c}
\tablecaption{Summary of Optically Detected Isolated Neutron Stars\label{tab:ins}}
\tablewidth{0pt}
\tablehead{
\colhead{Source} & \colhead{Period} & \colhead{$kT$\tablenotemark{a}} &
\colhead{$m_{V}$\tablenotemark{b}} & \colhead{$\log f_{X}/f_{V}$\tablenotemark{c}} &
\colhead{Optical\tablenotemark{d}} & \colhead{References}\\
 & \colhead{(s)} & \colhead{(eV)} & \colhead{(mag)} & & \colhead{Excess} \\
}
\startdata
\rxjw & \nodata & 61 & 25.8 & 4.4 &6 & 1,2,3,4  \\
\rxjk & 8.39 & 81 & 26.8 & 4.6 & 6 & 5,6\\
\rbs  & 10.31 & 91 & 28.7 & 5.0 & 5 & 7,8,9,10\\
\rxj  & \nodata & 92  & 27.1 & 4.4 & 14 & 11,12\\
\enddata
\tablenotetext{a}{Temperature of the best-fitting blackbody.}
\tablenotetext{b}{$V$-band Vega magnitude, either measured or interpolated.}
\tablenotetext{c}{Absorption-corrected bolometric X-ray-to-optical
  flux ratio, assuming that the X-ray emission is a blackbody.  The
  $V$-band flux is computed according to $f_{V}=10^{-(V+11.76)/2.5}\mbox{ ergs s}^{-1}\mbox{ cm}^{-2}$, following \citet*{bcp98}.}
\tablenotetext{d}{The ratio of the observed $V$-band flux to the
  extrapolated X-ray blackbody flux at 5500~\AA.}
\tablerefs{ 
1: \citet*{rgs02}; 2: \citet{bhn+02}; 3: \citet{dmd+02}; 4: \citet{vkk01}; 5:
\citet{hmb+97}; 6: \citet{kvkm+03}; 7: \citet{shs+99}; 8: \citet{hhss02}; 9:
\citet{kkvk02}; 10: \citet{haberl03}; 11: \citet{mhz+99}; 12: this work.}
\end{deluxetable}

The second scenario could occur if there were a significant contribution
to the optical emission from non-thermal emission.  Non-thermal
emission could arise if there were a substantial spin-down luminosity,
$\dot E$ (such as that seen for PSR~B0656+14; \citealt{kpz+01}).  In
this case, the high optical excess could indicate a large $\dot E$ for
\rxj.  While we cannot constrain the non-thermal emission from the
current photometry, it may be difficult to reproduce the thermal-like
spectrum observed in Figure~\ref{fig:nufnu} and to invoke significant
non-thermal flux.  One could appeal to light-element atmospheres for
\rxj, as they have high optical excesses over blackbodies
\citep{pztn96}.  However, these excesses are far too high (factors of
50--100, instead of 14) to fit our photometry, allowing us to reject
such models.

Future observations, such as higher-precision X-ray spectroscopy from
\chandra\ and \xmm\ (being analyzed), additional optical photometry,
and improved X-ray timing (in order to determine $P$ and eventually
$\dot E$ and the magnetic field $B$), should help to settle these
issues.  We are also searching for an H$\alpha$ nebula around \rxj\
(e.g.,\ \citealt{vkk01b}).  A single definitive measurement would be
the distance, which would determine the areas of the X-ray and optical
emission regions, but a parallax measurement with \hst\ would require
a significant investment of observing time.

\acknowledgements We thank an anonymous referee for helpful comments.
D.~L.~K.\ is supported by the Fannie and John Hertz
Foundation and S.~R.~K.\ by NSF and NASA.  Data presented herein were
based on observations made with the NASA/ESA Hubble Space Telescope,
obtained at the Space Telescope Science Institute, which is operated
by the Association of Universities for Research in Astronomy, Inc.,
under NASA contract NAS 5-26555.  Data presented herein were also
obtained at the W.~M.~Keck Observatory, which is operated as a
scientific partnership among the California Institute of Technology,
the University of California, and the National Aeronautics and Space
Administration.  The Guide Star Catalog-II is a joint project of the
Space Telescope Science Institute and the Osservatorio Astronomico di
Torino.  MIDAS is developed and maintained by the European Southern
Observatory.

\bibliographystyle{apj}


\end{document}